\documentclass[useAMS,usenatbib]{mn2e}

\usepackage{graphicx} 
\usepackage{epsfig}

\title[PKS 2123$-$463: a confirmed $\gamma$-ray blazar at high redshift]{PKS
  2123$-$463: a confirmed $\gamma$-ray blazar at high redshift}

\author[F. D'Ammando, A. Rau, P. Schady, et al.]{F. D'Ammando$^{1,2,3}$\thanks{E-mail: filippo.dammando@fisica.unipg.it},
  A. Rau$^{4}$, P. Schady$^{4}$, J. Finke$^{5}$, M. Orienti$^{6,3}$, J. Greiner$^{4}$, \newauthor
  D. A. Kann$^{7}$, R. Ojha$^{8,9}$, A. R. Foley$^{10}$, J. Stevens$^{11}$,
  J. M. Blanchard$^{12}$, \newauthor P. G. Edwards$^{13}$, M. Kadler$^{14,15}$, J. E. J. Lovell$^{12}$\\
$^{1}$Dip. di Fisica, Universit\`a degli Studi di Perugia, Via A. Pascoli, I-06123 Perugia, Italy \\
$^{2}$INFN - Sezione di Perugia, Via A. Pascoli, I-06123 Perugia, Italy \\
$^{3}$INAF - Istituto di Radioastronomia, Via Gobetti 101, I-40129 Bologna, Italy\\
$^{4}$Max-Planck-Institute f\"ur Extraterrestrische Physik,Giessenbachstra$\ss$e 1, 85748 Garching, Germany \\
$^{5}$U.S. Naval Research Laboratory, Code 7653, 4555 Overlook Ave. SW, Washington, DC 20375-5352, USA \\
$^{6}$Dip. di Astronomia, Universit\`a di Bologna, Via Ranzani 1, I-40127 Bologna, Italy \\ 
$^{7}$Th\"uringer Landessternwarte Tautenburg, Sternwarte 5, 07778 Tautenburg, Germany \\
$^{8}$NASA, Goddard Space Flight Center, Greenbelt, MD 20771, USA \\
$^{9}$IACS, Dept. of Physics, The Catholic University of America, 620 Michigan Ave., N.E., Washington, DC 20064, USA \\
$^{10}$SKA SA, Cape Town Office 3rd Floor, The Park, Park Road, Pinelands, Cape Town, South Africa\\
$^{11}$CSIRO Astronomy and Space Science, ATNF, Locked Bag 194, Narrabri NSW 2390, Australia\\
$^{12}$School of Mathematics \& Physics, Private Bag 37, University of Tasmania, Hobart TAS 7001, Australia \\
$^{13}$CSIRO Astronomy and Space Science, ATNF, PO Box 76, Epping NSW 1710, Australia\\
$^{14}$Institut f\"ur Theoretische Physik und Astrophysik, Universit\"at W\"urzburg,
97074 W\"urzburg, Germany \\
$^{15}$CRESST/NASA Goddard Space Flight Center, Greenbelt, MD 20771, USA}
\begin{document}

\date{Accepted. Received; in original form}


\maketitle


\begin{abstract}
The flat spectrum radio quasar (FSRQ) PKS 2123$-$463 was associated in the First {\em Fermi}-LAT source catalog with the $\gamma$-ray source 1FGL
J2126.1$-$4603, but when considering the full first two years of {\em Fermi} observations, no $\gamma$-ray source at a position consistent with this FSRQ
  was detected, and thus PKS 2123$-$463 was not reported in the Second {\em Fermi}-LAT source catalog. On 2011 December 14 a $\gamma$-ray source positionally consistent with PKS 2123$-$463 was detected in flaring activity by {\em Fermi}-LAT. This activity triggered radio-to-X-ray
observations by the {\em Swift}, GROND, ATCA, Ceduna, and KAT-7 observatories. Results of the localization of the $\gamma$-ray source over
41 months of {\em Fermi}-LAT operation are reported here in conjunction with
the results of the analysis of radio, optical, UV and X-ray data collected
soon after the $\gamma$-ray flare. 

\noindent The strict spatial association with the lower energy counterpart together with a simultaneous increase of the
activity in optical, UV, X-ray and $\gamma$-ray bands led to a firm
identification of the $\gamma$-ray source with PKS 2123$-$463. A new
photometric redshift has been estimated as $z = 1.46\pm0.05$ using GROND
and {\em Swift}/UVOT observations, in rough agreement with the disputed spectroscopic redshift of $z = 1.67$. We
  fit the broadband spectral energy distribution with a synchrotron/external
  Compton model. We find that a thermal disk component is necessary to explain
  the optical/UV emission detected by {\em Swift}/UVOT. This disk has a luminosity of $\sim 1.8\times10^{46}$ erg s$^{-1}$, and a fit to
the disk emission assuming a Schwarzschild (i.e., nonrotating) black hole gives a mass of $\sim
2\times10^{9}$ M$_{\odot}$. This is the first black hole mass estimate for this source.

\end{abstract}


\section{Introduction}

Blazars constitute the most extreme subclass of Active Galactic Nuclei (AGN),
characterized by the emission of strong non-thermal radiation across the
entire electromagnetic spectrum and in particular intense $\gamma$-ray
emission above 100 MeV. The typical observational properties of blazars
include irregular, rapid, high-amplitude variability, radio-core
dominance, apparent super-luminal motion, a flat radio spectrum, and high and variable polarization at
radio and optical frequencies. These features are interpreted as resulting
from the emission of high-energy particles accelerated within a relativistic
jet closely aligned with our line of sight and launched in the vicinity of the supermassive black hole harboured by the active galaxy \citep{blandford78,urry95}. 

Since the advent of the Energetic Gamma Ray Experiment Telescope (EGRET) on
the Compton Gamma-Ray Observatory (CGRO), blazars were known to dominate the extragalactic high-energy sky. However, EGRET
did not pinpoint the location of many sources with sufficient precision to
enable astronomers to associate them with known objects, leaving the legacy of
a large fraction of unidentified sources in $\gamma$ rays. The
point-spread function and sensitivity of the Large Area Telescope (LAT) on
board {\em Fermi} provides an unprecedented angular resolution at high
energies for localizing a large number of newly found $\gamma$-ray
emitting sources. Correlated variability observed at different frequencies can give important information for the identification of a $\gamma$-ray source with its low-energy counterpart.

On 2011 December 14 a $\gamma$-ray flare from a source positionally consistent with PKS
2123$-$463 was detected by {\em Fermi}-LAT \citep{orienti11}, triggering GROND
and {\em Swift} follow-up observations \citep{rau11, dammando11} that confirmed contemporaneous activity in the optical/UV as well as marginally in X-rays.

PKS 2123$-$463 is a bright radio quasar with a luminosity at 1.4 GHz
L$_{\rm 1.4\, GHz}$= (1.5$\pm$0.2)$\times$10$^{28}$ W/Hz  (assuming the
redshift estimated in this paper, $z$ = 1.46$\pm$0.05\footnote{The corresponding luminosity distance for $z$ = 1.46$\pm$0.05 is d$_L$ = 10.6$\pm$0.5 Gpc and 1 arcsec corresponds to a projected distance of 8.53$\pm$0.02 kpc.}; see Sect.~\ref{redshift}). On the basis of its
spectral index $\alpha_{r}\sim$0.4 (S($\nu$) $\propto \nu^{-\alpha_r}$)
between 408 MHz and 4.8 GHz, it was included in the CRATES catalog of flat
spectrum objects \citep{healey07}. ATCA observations performed almost
simultaneously at 4.8, 8.6 and 20 GHz during the Australia Telescope 20 GHz
(AT20G) survey indicated a flattening of the radio spectrum at high
frequencies to a spectral index of $\alpha_{r}\sim0.2$. Polarized emission has
been detected only at 4.8 GHz where the polarization is about 4\% of the total
intensity flux density \citep{massardi08}. A {\em Chandra} observation of PKS
2123$-$463 in March 2004 has shown the presence of a jet-like extended
structure in X-rays \citep{marshall11}. The redshift of PKS 2123$-$463 was
reported to be $z = 1.67$ \citep{savage81} based on an objective-prism
spectrum, subsequently questioned by \citet{jackson02} because two possible
redshifts (0.48 and 1.67) were given from observation of two lines in the
spectrum, and the motivation for the exclusion of the smaller redshift was
  not provided. This object was a member of the pre-{\em Fermi} launch Roma-BZCAT \citep{massaro09} catalog listing candidate $\gamma$-ray blazars detectable by {\em Fermi}-LAT but not in the CGRaBS catalog \citep{healey08}.

The flat spectrum radio quasar (FSRQ) PKS 2123$-$463 \citep[R.A.:21h26m30.7042s,
Dec.:$-$46d05m47.892s, J2000;][]{fey06} was associated in the First {\em
  Fermi}-LAT source catalog \citep[1FGL, 2008 August 4 -- 2009 July 4;][]{abdo10a} with the $\gamma$-ray source 1FGL
J2126.1$-$4603, while no association with a $\gamma$-ray source was present in
the Second {\em Fermi}-LAT source catalog \citep[2FGL, 2008 August 4 -- 2010
August 4;][]{nolan12}, although a
$\gamma$-ray source, 2FGL J2125.0$-$4632 with a radius of the 95\% source
location confidence region of 0.17$^{\circ}$, at 0.52$^{\circ}$ from the radio
position of PKS 2123$-$463 was reported. Taking into consideration the high
variability of blazars, the flux of PKS 2123$-$463 could have decreased in the second year of {\em
  Fermi}-LAT operation. In addition, in the construction of the 2FGL catalog,
PKS 2123$-$463 was split into more than one candidate source seed \citep[see
Section 4.2 of][for details]{nolan12}. These two factors may have led to the lack of its association with a $\gamma$-ray source. 

In this paper we present the localization over 41 months of {\em Fermi}-LAT
data of a $\gamma$-ray source associated with the FSRQ PKS 2123$-$463. The correlated variability observed in optical,
UV, X-ray and $\gamma$ rays confirms the identification.
In addition a new estimation of the redshift of the source by means of the fit of
simultaneous GROND and {\em Swift}/UVOT data collected soon after the
$\gamma$-ray flare is presented. The paper is organized as follows: in
Section 2 we report the LAT data analyses and results. In Sections 3 and 4 we
report the results of the {\em Swift} and GROND data analysis, respectively. Radio data collected by the ATCA, Ceduna and KAT-7 telescopes are
presented in Section 5. Section 6 presents an estimation of the redshift of
the source. In Section 7 we discuss the modeling of the overall SED and draw
our conclusions. Throughout the paper a $\Lambda$--CDM cosmology with H$_0$ = 71 km s$^{-1}$ Mpc$^{-1}$,
$\Omega_{\Lambda}$ = 0.73, and $\Omega_{m}$ = 0.27 is adopted.

\section{{\em Fermi}-LAT Data: Selection and Analysis}
\label{FermiData}

The {\em Fermi}-LAT is a $\gamma$-ray telescope operating from $20$\,MeV to
$>300$\,GeV. It has a large peak effective area ($\sim$ $8000$\,cm$^2$ for $1$\,GeV photons), an
energy resolution of typically $\sim 10\%$, and
a field of view of about $ 2.4$ \,sr with an angular resolution ($68\%$ containment angle) better than 1$^{\circ}$ for
energies above $1$\,GeV. Further details about the {\em Fermi}-LAT are given by \citet{atwood09}.

\begin{figure}
\centering
\includegraphics[width=7cm]{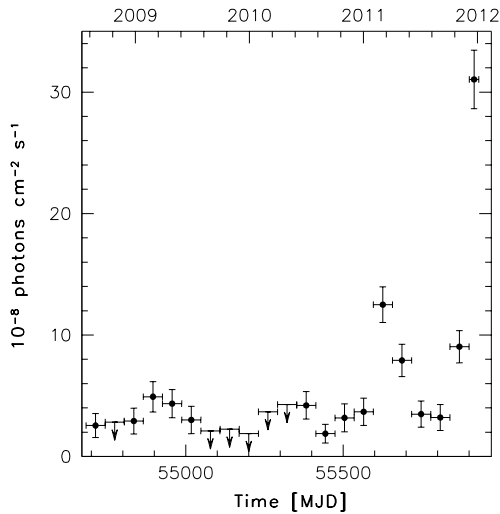}
\caption{2-month integrated flux (E $>$ 100 MeV) light curve of PKS 2123$-$463
  obtained from 2008 August 4 to 2012 January 4 (MJD 54682--55930). The last bin covers 1 month
  of observation. Arrows refer to 2-$\sigma$ upper limits on the source flux. Upper limits are computed when TS $<$ 10.}
\label{Fig1}
\end{figure}

The {\em Fermi}-LAT data reported in this paper were collected over the first
41 months of {\em Fermi} operation, from 2008 August 4 (MJD 54682) to 2012
January 4 (MJD 55930). During this time the LAT instrument operated almost entirely in survey
mode. The spectral analysis was performed with the \texttt{ScienceTools} software
package version v9r23p1. The {\em Fermi}-LAT data were extracted from a circular Region
of Interest (RoI) with a $15^{\circ}$ radius centred at the radio location of PKS
2123$-$463. Only events belonging to the `Source' class were used. The time
intervals when the rocking angle of the LAT was greater than 52$^{\circ}$ were
rejected. In addition, a cut on the zenith angle ($< 100^{\circ}$) was also applied to reduce contamination from the Earth limb $\gamma$ rays, which are produced by
cosmic rays interacting with the upper atmosphere. The spectral analyses (from
which we derived spectral fits and photon fluxes) were performed with the post-launch instrument  response functions (IRFs) \texttt{P7SOURCE\_V6}
using an unbinned maximum likelihood method implemented in the Science tool \texttt{gtlike}.

\begin{figure}
\centering
\includegraphics[width=7cm]{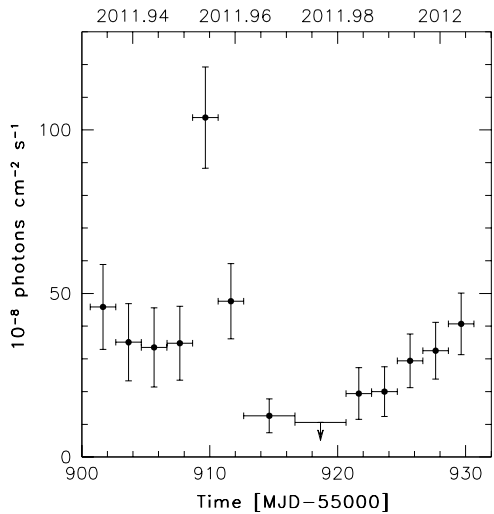}
\caption{Integrated flux (E $>$ 100 MeV) light curve of PKS 2123$-$463
  obtained from 2011 December 4 to 2012 January 4 (MJD 55899--55930) with 2-day or 4-day time bins. Arrows refer to 2-$\sigma$
  upper limits on the source flux. Upper limits are computed when TS $<$ 10.}
\label{Fig2}
\end{figure}

The background model used to extract the $\gamma$-ray signal includes a
Galactic diffuse emission component and an isotropic component. The model that
we adopted for the Galactic component, the same as used for the 2FGL catalog, is given by the file gal\_2yearp7v6\_v0.fits, and the isotropic component, which is the sum of the extragalactic diffuse
emission and the residual charged particle background, is parametrized by the
file
iso\_p7v6source.txt\footnote{http://fermi.gsfc.nasa.gov/ssc/data/access/lat/Background\\Models.html}.
 The normalizations of both components in the background model were allowed to vary freely during the spectral point fitting. 

We examine the significance of the $\gamma$-ray signal from the sources using
the Test Statistic (TS) based on the likelihood ratio test. The Test
Statistic, defined as TS = 2$\Delta$log(likelihood) between models with and without the
source, is a measure of the probability of having a $\gamma$-ray source at the
localization specified, which compares models whose parameters have been
adjusted to maximize the likelihood of the data given the model
\citep{mattox96}. The source model used in \texttt{gtlike} includes all the point sources from the 2FGL that fall
within $20^{\circ}$ of PKS 2123$-$463. The spectra of those sources were
parametrized by power-law functions, except for 2FGL J2056.2$-$4715 for which
we used a log-parabola, and 2FGL J2124.6$-$3357 and 2FGL J2241.7$-$5236 for which we used an exponentially
cut-off power-law in their spectral modeling as in the 2FGL catalog. We removed
from the model the sources having TS $<$ 25 and/or fluxes (0.1--100 GeV) below 1.0$\times$10$^{-8}$ photons cm$^{-1}$ s$^{-1}$ over 41
months and repeated the fit. We tested whether two distinct $\gamma$-ray
  sources (one at the radio position of PKS 2123$-$463 and one at the
  $\gamma$-ray position of 2FGL J2125.0$-$4632) are detected simultaneously by
  {\em Fermi} with TS $\geq$ 25 over 41 months of observations. In this
  scenario, the fit yields a TS of 553 for PKS 2123$-$463 and a TS of 14 for
  2FGL J2125.0$-$4632. We thus conclude that PKS 2123$-$463 and 2FGL
  J2125.0$-$4632 are the same source and we maintain only PKS 2123$-$463 in
  the model. Thus a final fitting procedure has been performed with all
sources within 10$^{\circ}$ of PKS 2123$-$463 included with
the normalization factors and the photon indices left as free
parameters. For the sources located between 10$^{\circ}$ and 15$^{\circ}$ we kept
the normalization and the photon index fixed to the values obtained in the
previous fitting procedure. The RoI model includes also sources falling
between 15$^{\circ}$ and 20$^{\circ}$ from the target source, which can contribute to the total counts observed in the RoI
due to the energy-dependent size of the point spread function of the instrument. For these additional sources, normalizations and indices were fixed to the values of the 2FGL catalog during all steps of the fitting procedure.

The $\gamma$-ray point source localization using the \texttt{gtfindsrc}
tool over the photons extracted during the period 2008 August 4 -- 2012
January 4 results in R.A. = 321.609$^{\circ}$, Dec. = $-$46.076$^{\circ}$ (J2000), with a
95$\%$ error circle radius of 0.047$^{\circ}$, at an angular separation of 0.024$^{\circ}$ from the radio
position of PKS 2123$-$463 (R.A. = 321.628$^{\circ}$, Dec. = $-$46.097$^{\circ}$, J2000). This implies a
strict spatial association with PKS 2123$-$463. 
The fit with a power-law (PL) model, dN/dE $\propto$ (E/E$_{0}$)$^{-\Gamma}$, to the data integrated over 41
months of {\em Fermi} operation in the 0.1--100 GeV energy range results in a TS = 557, with an
integrated average flux of (4.76 $\pm$ 0.35) $\times$10$^{-8}$ photons
cm$^{-2}$ s$^{-1}$, and a photon index $\Gamma$ = 2.45 $\pm$ 0.05 (with E$_0$ fixed to 670 MeV). In order to
test for curvature in the $\gamma$-ray spectrum of PKS 2123$-$463 we used as an
alternative spectral model the log-parabola (LP), dN/dE $\propto$ (E/E$_{0}$)$^{-\alpha-\beta \,\rm  log(E/E_0)}$
\citep{landau86, massaro04}, where the parameter $\alpha$ is  the spectral
slope at the energy E$_0$ and the parameter $\beta$ measures the curvature
around the peak. We fixed the reference energy E$_0$ to 300 MeV. The fit with
a log-parabola results in a TS = 554, with spectral parameters $\alpha$ = 2.35
$\pm$ 0.09, and $\beta$ = 0.06 $\pm$ 0.04. Applying a likelihood ratio test to check the PL model (null hypothesis) against the LP model (alternative
hypothesis), the power-law spectral model is favoured, indicating that no
significant curvature was observed in the average $\gamma$-ray spectrum. 

Figure~\ref{Fig1} shows the $\gamma$-ray light curve of the first 41 months of
{\em Fermi} observations of PKS 2123$-$463 built using 2-month time bins, with
the exception of the final (1-month) data point. For each time bin the photon index was frozen to the value
resulting from the likelihood analysis over the entire period. If TS $<$ 10
2-$\sigma$ upper limits were calculated instead. The systematic uncertainty in the flux is energy dependent: it amounts to $10\%$
at 100 MeV, decreasing to $5\%$ at 560 MeV, and increasing to $10\%$ above 10
GeV \citep{ackermann12}. 

As shown in Fig.~\ref{Fig1} PKS 2123$-$463 was in a low $\gamma$-ray state
(0.1--100 GeV flux $<$ 5$\times$10$^{-8}$ ph cm$^{-2}$ s$^{-1}$) for the first
2.5 years. In particular the source was not detected with 2-month time bin during the period 2009 August--2010 May. A first increase of the $\gamma$-ray flux was observed in
2011 February-March, and subsequently a flaring activity in 2011 December. We
show a light curve focused on the period of high activity (2011 December
4--2012 January 4; MJD 55899--55930) with 2-day or 4-day time bins
(Fig.~\ref{Fig2}). The peak of the emission was observed between December 13 15:43 UT and December 14 15:43
UT, with a flux of (128 $\pm$ 23) $\times$10$^{-8}$ photons cm$^{-2}$
s$^{-1}$ in the 0.1--100 GeV energy range, a factor of $\sim$ 25 higher with
respect to the average $\gamma$-ray flux observed during 2008-2011. The
corresponding observed isotropic $\gamma$-ray luminosity peak in the 0.1--100
GeV energy range is 8.9$\pm$0.8 $\times$ 10$^{48}$ erg s$^{-1}$ (assuming $z$ = 1.46$\pm$0.05, see
Sect.~\ref{redshift}), comparable to the values reached by the most powerful FSRQs in flaring activity \citep[e.g.][]{orienti12,ackermann10}. 
Leaving the photon index free to vary during the period of high activity and
$E_0$ fixed to 670 MeV the fit results in a photon index $\Gamma$ = 2.26 $\pm$ 0.06, showing a moderate harder-when-brighter behaviour already observed in other FSRQs \citep{abdo10b}.
Replacing in the same period the PL with a LP, fixing the
reference energy E$_0$ to 300 MeV, we obtain spectral parameters $\alpha$ = 2.09 $\pm$ 0.13, and $\beta$ = 0.10 $\pm$ 0.04. We used a likelihood ratio
test to check the PL model (null hypothesis) against the LP model (alternative
hypothesis). These values may be compared, following \citet{nolan12}, by evaluating the curvature Test Statistic TS$_{curve}$ =
2(log(like)$_{LP}$-log(like)$_{PL}$)=3 ($\sim$1.7-$\sigma$); thus no
significant curvature was observed in the $\gamma$-ray spectrum also during the high activity period.

Finally, using the \texttt{gtsrcprob} tool, which calculates the
  probabilities an event belongs to each of the sources in the model, we estimated that the highest-energy photon emitted by PKS 2123$-$463 was observed with a
probability of 87\% at distance of 0.35$^{\circ}$ from the source with an
energy of 92 GeV. For a single high-energy photon ($\geq$ 1 GeV), sky
  location can only be determined within $\sim$ 0.8$^{\circ}$ (95\% confidence radius).

\section{Swift Data: Analysis and Results}
\label{SwiftData}

The {\em Swift} satellite \citep{gehrels04} performed two observations of
PKS 2123$-$463 in 2011 December triggered by the $\gamma$-ray flare detected
by {\em Fermi}-LAT. As a comparison we also analysed two additional observations
performed during 2010--2011. The observations were performed with all three on-board instruments: the X-ray Telescope \citep[XRT;][0.2--10.0 keV]{burrows05},
the UltraViolet Optical Telescope \citep[UVOT;][170--600 nm]{roming05} and the Burst Alert Telescope \citep[BAT;][15--150 keV]{barthelmy05}. 

The hard X-ray flux of this source is below the sensitivity of the BAT instrument for the short exposure time of these observations. Moreover, the source was not
present in the {\em Swift} BAT 58-month hard X-ray catalog \citep{baumgartner10} and the 54-month Palermo BAT catalog \citep{cusumano10}.

\begin{table*}
\caption{Log and fitting results of {\em Swift}/XRT observations of PKS
  2123$-$463 using a power-law model with $N_{\rm H}$ fixed to Galactic
absorption.}
\label{XRT}      
\centering                          
\begin{tabular}{c c c c c}       
\hline\hline 
MJD  &  UT Date & Exposure Time & $\Gamma$ &  Flux (0.3--10 keV)$^{a}$ \\             
     &       & [sec]   &  & [$\times$10$^{-12}$ erg cm$^{-2}$ s$^{-1}$] \\    
\hline                        
55467 & 2010-09-28 & 2712 & 1.47$\pm$0.30 & 1.31$\pm$0.46  \\
55745 & 2011-07-03 & 2912 & 1.61$\pm$0.28 & 1.20$\pm$0.32 \\
55910 & 2011-12-15 & 4947 & 1.68$\pm$0.18 & 1.90$\pm$0.34 \\
55914 & 2011-12-19 & 5521 & 1.26$\pm$0.18 & 1.83$\pm$0.27  \\
\hline                                  
\end{tabular}
\\
$^{a}$: Observed flux.
\end{table*}

\begin{table*}
\caption{UVOT Photometry of PKS 2123$-$463.}
\begin{tabular}{lccccccc}
\hline\hline
UT Date &\multicolumn{6}{c}{AB Magnitude$^a$} \\
& $uvw2$ & $uvm2$ & $uvw1$ & $u$ & $b$ & $v$ \\
\hline
2010-09-28  18:29-23:30 &  $19.78\pm0.09$ & $19.34\pm0.10$ & 
$19.17\pm0.10$ & $18.66\pm0.09$ & $18.61\pm0.14$ & $18.59\pm0.20$\\
2011-07-03  17:15-23:29 &  $19.83\pm0.20$ & $19.28\pm0.14$ & 
$18.75\pm0.11$ & $18.47\pm0.12$ & $18.81\pm0.25$ & $18.21\pm0.26$\\
2011-12-15  20:45-01:25 &  $18.97\pm0.06$ & $18.58\pm0.06$ & 
$18.24\pm0.05$ & $17.95\pm0.06$ & - & $17.64\pm0.06$\\
2011-12-19  04:20-04:39 &  $19.17\pm0.12$ & $18.80\pm0.12$ & 
$18.51\pm0.09$ & $18.04\pm0.07$ & $17.77\pm0.07$ & $17.82\pm0.15$\\
\hline
\end{tabular}
\\
$^a$: Corrected for Galactic foreground reddening of E$_{\rm 
B-V}=0.03$\,mag \citep{schlegel98}.\\
\label{tab:uvot}
\end{table*}

\subsection{Swift/XRT}

The XRT data were processed with standard procedures (\texttt{xrtpipeline v0.12.6}), filtering, and screening criteria by using the \texttt{HEAsoft} package
(v6.11)\footnote{http://heasarc.nasa.gov/lheasoft/}. The data were collected in photon counting mode in all observations, and only XRT event grades 0--12 were selected. The source count rate
was low ($<$ 0.5 cnt s$^{-1}$); thus pile-up correction was not
required. Source events were extracted from a circular region with a radius of
20 pixels (1 pixel $\sim$ 2\farcs36), while background events were extracted
from a circular region with a radius of 50 pixels away from the source region. Ancillary response files were generated
with \texttt{xrtmkarf}, and account for different extraction regions,
vignetting and point spread function (PSF) corrections. We used the spectral redistribution matrices
v013 in the Calibration database maintained by HEASARC. 

Considering the low number of photons collected ($<$ 200 counts) the spectra
were rebinned with a minimum of 1 count per bin and the Cash statistic
\citep{cash79} was applied. We fitted the spectrum with an absorbed power-law using
the photoelectric absorption model \texttt{tbabs} \citep{wilms00}, with a neutral
hydrogen column fixed to its Galactic value \citep[2.34$\times$10$^{20}$
cm$^{-2}$;][]{kalberla05}. The fit results are reported in
Table~\ref{XRT} and Fig.~\ref{correlated}. 

A marginal increase (1.5 $\sigma$) of the X-ray flux with respect to the previous {\em
  Swift}/XRT observations was observed on 2011 December 15, soon after the
$\gamma$-ray flare. The source remained at a similar flux level in X-rays on 2011 December 19.

\subsection{Swift/UVOT}

UVOT UV/optical imaging was obtained during all four {\em Swift} observations of PKS~2123$-$463. The photometry was carried out on pipeline processed sky
images downloaded from the {\em Swift} data
center\footnote{http://www.swift.ac.uk/swift\_portal}, following the standard
UVOT procedure \citep{Poole:2008qy}. Source photometric measurements were extracted from the UVOT imaging  data using the tool {\sc uvotmaghist} (v1.1) with a circular source extraction region that ranged from
3\farcs5-5\arcsec radius to maximize the signal-to-noise. In order to remain compatible with the effective area calibrations, which are based on
5\arcsec aperture photometry \citep{Poole:2008qy}, an aperture correction was applied  where necessary.  This correction was at maximum  5--6\,\%
of the flux, depending on the filter. The UVOT photometry is presented in
Table~\ref{tab:uvot} and Fig.~\ref{correlated}. Contemporaneous UVOT observations in 2011 December 15 found
PKS 2123$-$463 about 0.5 mag brighter in $v$-, $u$-, and $w1$-band, 0.7 mag in
$m2$-band, and about 0.8 mag in $w2$-band compared to the UVOT observation performed on 2011 July 3.   

\section{GROND data}

On 2011 December 18, at 01:21 UT, PKS~2123$-$463 was observed with the
Gamma-ray Optical/Near-Infrared Detector \citep[GROND;][]{greiner08}
mounted on the MPG/ESO 2.2-m telescope at  La Silla, Chile. Preliminary
results  have been reported in \citet{rau11}. GROND is a 7-channel imager
that  observes in four optical and three near-IR channels simultaneously. The
data were reduced and analysed with the  standard tools and methods described in
\citet{Kruhler:2008aa}. Calibration was performed against an SDSS standard star field ($g^{\prime}r^{\prime}i^{\prime}z^{\prime}$)  and  against
selected 2MASS stars \citep{Skrutskie:2006wd} ($J H  K_S$). This resulted in 1-$\sigma$ accuracies of 0.05\,mag ($g^\prime  r^\prime i^\prime  z^\prime$, J, H), and  0.07\,mag ($K_S$).
All magnitudes are corrected for Galactic foreground extinction of E$_{\rm B-V}=0.030$\,mag
\citep{schlegel98} and are summarised in Table~\ref{tab:grond}.

\begin{table*}
\caption{GROND Photometry of PKS 2123$-$463.}
\begin{tabular}{lccccccc}
\hline\hline
UT Date&\multicolumn{7}{c}{AB Magnitude$^{a}$} \\
& $g^\prime$ & $r^\prime$ & $i^\prime$ & $z^\prime$ & $J$ & $H$ & $K_s$\\
\hline
2011-12-18 01:21 & $17.78\pm0.05$ & $17.52\pm0.05$ & $17.43\pm0.05$
& $17.24\pm0.05$ & $16.75\pm0.06$ & $16.53\pm0.06$ & $16.24\pm0.09$ \\
\hline
\end{tabular}
\\
$^a$: Corrected for Galactic foreground reddening of E$_{\rm 
B-V}=0.03$\,mag \citep{schlegel98}.\\
\label{tab:grond}
\end{table*}

\section{Radio data}

PKS\,2123$-$463 is monitored by the TANAMI program \citep[Tracking Active
galactic Nuclei with Austral Milliarcsecond Interferometry;][]{ojha10} at a
number of radio frequencies and resolutions with the Australia Telescope Compact
Array (ATCA) and the Ceduna facilities. ATCA is observing this source every few weeks with ``snapshot''
observations at frequencies from 5.5\,GHz through 40\,GHz where each frequency is
the centre of a 2 GHz wide band and the fluxes are calibrated against the ATCA
primary flux calibrator PKS 1934$-$638 \citep{stevens12}. Each flux density
has a 1-$\sigma$ uncertainty of 0.01 Jy. The Ceduna radio telescope in South
Australia is monitoring PKS\,2123$-$463 at 6.7\,GHz. Each flux density
has a 1-$\sigma$ uncertainty of 0.3 Jy \citep{mcculloch05}. No sign of increased activity of the flux density was detected between 40 and 5.5
GHz before the $\gamma$-ray flare (see Fig.~\ref{correlated}).

\begin{table}
\caption{ATCA (5.5, 9, 17, 19, 38, and 40 GHz) and Ceduna (6.7 GHz) observations of PKS 2123$-$463 in 2011.}
\begin{tabular}{ccc}
\hline\hline
UT Date & Frequency (GHz) &  Flux (Jy) \\
\hline
2011-05-17 &      5.5	& 0.86 \\
	   &      9.0   & 0.77 \\
	   &     17.0   & 0.67 \\
	   &     19.0   & 0.64 \\
	   &     38.0   & 0.45 \\
           &     40.0	& 0.44 \\
\hline
2011-10-14 &     5.5    & 0.86 \\
	   &	 9.0    & 0.76 \\
           &    17.0    & 0.60 \\
           &    19.0    & 0.56 \\
           &    38.0    & 0.42 \\
           &    40.0    & 0.41 \\
\hline
2011-11-08 &    5.5  & 0.85 \\    
           &    9.0  & 0.71 \\
           &   17.0  & 0.57 \\
           &   19.0  & 0.53 \\
           &   38.0  & 0.43 \\
           &   40.0  & 0.43 \\
\hline
2011-12-19 &   6.7   & 0.81 \\
\hline
2012-02-18 &   6.7   & 0.78 \\
\hline
\end{tabular}
\label{tanami:tab}
\end{table}

In addition observations of PKS 2123$-$463 were made after the $\gamma$-ray outburst using the KAT-7 array (the prototype for MeerKAT) in the Karoo. Since
the array is still undergoing commissioning tests some uncertainties remained
about the absolute calibration scale. To minimize these uncertainties the observations were taken at similar times of
days (hence Local Sidereal Time range) in late 2011 December to early 2012 January, with 4-5~hr durations. The very first observation (December 21) failed because the online fringe stopping was not working properly, but as the array is very
small (longest baseline 200-m) and the frequency low enough (1.822 GHz) the delay tracking corrections could be done offline with only marginal loss in
signal to noise ratio. Subsequent observations were done in this mode. There
were also different problems with various receivers over this time so {\em
  uv}-coverage and maps were not directly comparable. The bandpass and gain
solutions (on PKS 1934$-$638) were good enough and
checks against a nearby phase reference source (PKS 2134$-$470) gave
sufficient confidence in the visibilities and phases to make simple maps of
the inner 0.5$^{\circ}$ of the primary beam and verify that no sources
stronger than 100 mJy were present. 

Assuming a flux density for PKS 1934$-$638 of 13.6 Jy at 1.822 GHz (as
interpolated from ATCA models) the measured flux density of PKS 2123$-$463 is 1.13 Jy on 2011 December 24, 26, 29, and 2012 January 7 and
  13. The uncertainties on the absolute flux density scale is 5\%. As a comparison we
extrapolate the flux density at 1.82 GHz using the values at 2.7, 0.8, and 0.4
GHz from the Parkes Catalogue (1990) and the Molonglo Sky Survey
(2008). Assuming the spectral index 0.4 derived between 2.7 and 0.4 GHz, we
obtain a flux density at 1.82 GHz of 1.0 Jy, indicating that no significant variation was observed during the $\gamma$-ray flaring activity at this frequency.

\section{New photometric redshift estimation}
\label{redshift}

\begin{figure}
\centering
\includegraphics[width=0.45\textwidth]{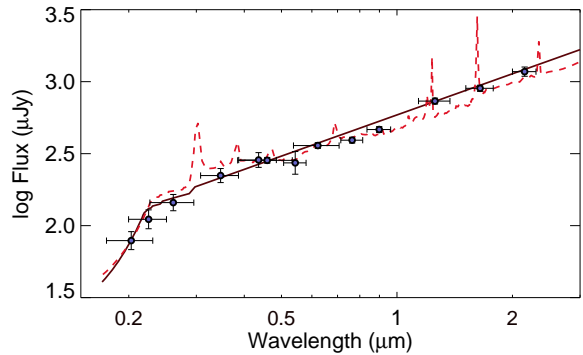}
\caption{UV-near-IR SED composed of {\em Swift}/UVOT observations
   obtained on 2011 December 19 and GROND data taken on 2011 December 18. The
simple power-law model (solid line) and the AGN/galaxy hybrid
template (dashed line) suggest a photometric redshift of
$z\approx1.45$ (see text for details).}
\label{fig:photoz}
\end{figure}

\begin{figure}
\centering
\includegraphics[width=0.45\textwidth]{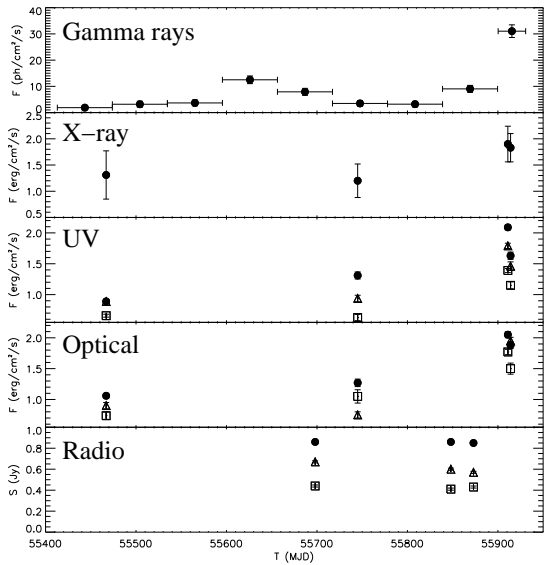}
\caption{Light curves of PKS 2123$-$463 collected in $\gamma$ rays by {\em
    Fermi}-LAT (0.1--100 GeV flux in units of 10$^{-8}$ ph cm$^{-2}$ s$^{-1}$), in X-rays by {\em Swift}/XRT (flux in units of 10$^{12}$ erg
  cm$^{-2}$ s$^{-1}$), in UV (filters w1: empty squares, m2: empty triangles,
  w2: filled circles; flux in units of 10$^{12}$ erg cm$^{-2}$ s$^{-1}$), and in optical
  (filters u: empty squares, b: empty triangles, v: filled circles; flux in units of
  10$^{13}$ erg cm$^{-2}$ s$^{-1}$) by {\em Swift}/UVOT, and radio by ATCA
  (empty squares: 40 GHz, empty triangles: 17 GHz, filled circle: 5.5 GHz) between 2010 June and 2011 December.}
\label{correlated}
\end{figure}

The redshift of PKS 2123$-$463 has not been established convincingly
yet. Following the method described in \citet{rau12} and applied to a sample
of 103 {\em Fermi}-LAT blazars, we combined the GROND photometry with the UVOT
photometry from 2011 December 19 in order to construct a 13-band spectral energy
distribution (SED). In order to account for source-intrinsic variability
between the GROND and UVOT observing epochs, the GROND photometry was corrected 
by $-0.05$\,mag in all bands \citep[see][for a detailed description of the method]{rau12}. The resulting SED, corrected for the 
Galactic foreground reddening of E$_{\rm B-V}=0.03$\,mag \citep{schlegel98},
has been fit with a set of power-law models as well as with hybrid templates
built from normal galaxies and AGN \citep{salvato09, salvato11} using the Le PHARE code \citep{arnouts99,
  ilbert06}. The best-fitting redshifts for both template libraries are in good agreement (see
Fig.~\ref{fig:photoz}). For the  power-law model we obtain a 99\% confidence redshift of $z = 1.37^{+0.13}_{-0.21}$ ($\chi^2=17$, P$_{\rm z}=94.6$ \footnote{The best fit templates and redshifts were selected  on  the basis of a simple $\chi^2$ method. P$_{\rm z}$ is the integral of the probability distribution function $\int f(z)  dz$ at $z_{\rm phot} \pm0.1(1+z_{\rm phot})$, which describes the probability that the  redshift of a  source is within $0.1(1+z)$ of the best fit value.}) and for the hybrid models we find $z = 1.46\pm0.05$ ($\chi^2=23$, P$_{\rm z}=99.2$). 
This is in rough agreement with the initial spectroscopic redshift of $z$ = 1.67 from \citet{savage81}. 

\section{SED modeling and conclusions}

Beyond the excellent spatial association obtained with the 41-month {\em
  Fermi}-LAT data set, the most secure and distinctive signature for firm
identification of the $\gamma$-ray source detected by {\em Fermi}-LAT with the
blazar PKS 2123$-$463 is the simultaneous increase observed in the
$\gamma$-ray, X-ray, and optical-UV bands (see Fig.~\ref{correlated}, Sections~\ref{FermiData} and
\ref{SwiftData}). 

In order to investigate the physical properties of the source we have built a
simultaneous SED of the flaring state of PKS 2123$-$463. The {\em Fermi}-LAT spectrum was
built with data from observations centred on 2011 December 10 to 19 (MJD 55905--55916). In
addition we included in the SED the GROND and {\em Swift} (UVOT and XRT) data collected on 2011
December 18 and December 19, respectively. Here, the flux suppression, in particular in the UV bands, was corrected assuming a power-law spectral shape consistent with the optical-near-IR measurements ($F_\lambda\propto\lambda^{0.8}$).
The data from ATCA and KAT-7 collected on December 19 and 24, respectively, provided information about
the radio part of the spectrum. The flare centred on $\sim$\ MJD 55909 had a
variability timescale of $\sim 2$\ days, which constrains the size of the
emitting region during the flare to R$^{\prime}_b \la 2.1\,\times 10^{16} (\delta_D / 20)$\ cm.

A ``blue bump'' accretion disc component is clearly visible in the optical/UV
data. We modeled these data with a combination of a nonthermal synchrotron
component and a Shakura-Sunyaev disc component \citep{shakura73}. A fit to the disc
component allows us to get a rough estimate of the black hole mass of
$M_{BH}\approx2\times10^9 M_{\odot}$.  This is the first black hole mass
estimate for this source.  It is consistent with black hole estimates for other high-$z$ FSRQs, obtained in a similar way by \citet{ghisellini11}. Our
mass estimate for PKS~2123$-$463 follows \citet{ghisellini11} and assumes that the innermost stable circular orbit (ISCO) is $R_{disc,ISCO}=6 R_g$, as one
would expect for a Schwarzschild (i.e., nonrotating) black hole.  If the jet
is produced by the Blandford-Znajek \citep{blandford77} or Blandford-Payne
\citep{blandford82} mechanisms this requires a nonzero black hole spin, and
one expects $R_{disc,ISCO}$ to be smaller or larger, depending on whether the
spin is retrograde or prograde \citep[e.g.,][]{garofalo10}. Due to the uncertainty in spin, this mass estimate can be considered to have considerable uncertainty.

We modelled the portion of the SED from X-rays to $\gamma$ rays assuming emission from a relativistic jet with mechanisms of synchrotron self-Compton (SSC),
and Compton-scattering of a dust torus external to the jet (EC-dust). The
description of the model can be found in \citet{finke08} and \citet{dermer09}. The synchrotron component considered is self-absorbed below
$\sim 10^{11}$\ Hz. Correlations of $\gamma$-ray and optical flares with radio light curves and rotations
of optical polarisation angles in low-synchrotron-peaked blazars seem
to indicate the $\gamma$-ray/optical emitting region is outside the
broad line region (BLR), where the dust torus is the likely seed
photon source \citep[e.g.,][]{marscher10}. From the disc luminosity obtained
by the fit of the blue bump, $L_{disc}=1.8\times10^{46}$\ erg s$^{-1}$,
we estimate an associated BLR radius $R_{BLR}=4.4\times10^{17}$\ cm, based on
the relation between disc luminosity and $R_{BLR}$ determined from
reverberation mapping campaigns \citep[e.g.,][]{kaspi05, ghisellini08}.
To minimize the scattered BLR contribution, we placed the emitting
region at $r>R_{BLR}$.  Here the primary seed photon source is the
dust torus, which was simulated as a one-dimensional ring with radius
$R_{dust}$ aligned orthogonal to the jet, emitting as a blackbody with
temperature $T_{dust}$ and luminosity $L_{dust}$.

\begin{figure} 
\centering
\vspace{3.5mm}
\hspace{-3.5mm}
\includegraphics[width=7.5cm]{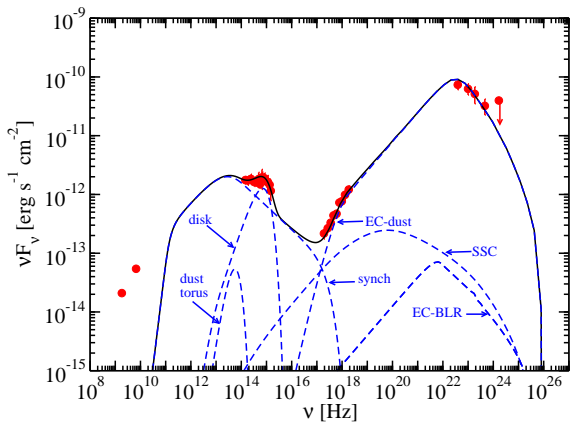}
\caption{Spectral energy distribution data (circles and squares) and model fit (solid curve) of PKS 2123$-$463 with the model components shown as dashed
  curves. The data points were collected by GROND (2011 December 18), {\em Swift} (UVOT and XRT; 2011 December 19), and {\em Fermi}-LAT (2011
  December 10-19), together with radio data from ATCA (2011 December 19) and
  KAT-7 (2011 December 24).}
\label{2123sed}
\end{figure}

The model fit to the broadband SED can be seen in Fig.~\ref{2123sed}
and the parameters can be found in Table~\ref{table_fit}.  A
description of the parameters can be found in \citet{dermer09}. The
dust parameters were chosen so that $R_{dust}$ is roughly consistent
with the sublimation radius \citep{nenkova08}. The EC-BLR component
was calculated assuming the seed photons are from H$\alpha$ and have a
luminosity of $0.1 L_{disc}$.

The electron distribution used, a broken power-law with index
$p_1=2.0$ below the break at $\gamma^{\prime}_{brk}$ and $p_2=3.8$ for
$\gamma^{\prime}_{brk}< \gamma^{\prime}$, is consistent with particles
injected with index $2.8$, and emission taking place in the {\em fast
  cooling regime} \citep[e.g.,][]{boett02}. That is, particles are
injected between $\gamma^{\prime}_{brk}$ and $\gamma^{\prime}_{max}$,
with a cooling electron Lorentz factor
$$
\gamma_{cool}\equiv \frac{3 m_ec^2}{4 c \sigma_T t_{esc} u_{tot} }\
$$ 
where $u_{tot}$ is the total energy density in the blob frame, which
in the case of our model fit is dominated by the external energy
density.  In this case $\gamma_{cool}$ is associated with
$\gamma_{min}$, since it is in the fast cooling regime.  
Also note that in this model fit the magnetic field and electrons are nearly in
equipartition. Jet powers were calculated assuming a two-sided jet.

The Compton dominance for PKS 2123$-$463, i.e., the ratio of the peak
luminosities of the Compton and synchrotron components, is $\approx50$, which is a rather standard value for powerful blazars
\citep[e.g.,][]{ghisellini11}. A large disc luminosity was estimated from the UVOT data, as expected for a powerful FSRQ, with a
L$_{disc}$/L$_{Edd}$=0.2 in agreement with the blazar divide proposed by \citet{ghisellini09} as a result of the changing of the accretion mode. 

Variability is common in $\gamma$-ray blazars and provides a powerful tool to associate them definitively with objects known at other wavelengths and to study
the emission mechanisms at work. The combination of deep and fairly uniform exposure over $\sim$3~hr, very good angular resolution, and stable response of
the {\em Fermi}-LAT is producing the most sensitive, best-resolved survey of
the $\gamma$-ray sky. On the other hand, those cases where there is a decrease in the activity, and thus of the significance of detection, can lead to a more
complex identification process for a $\gamma$-ray source. When combined with
simultaneous ground- and space-based multifrequency observations, the {\em Fermi}-LAT achieves its full capability for the identification of the
$\gamma$-ray sources with counterparts at lower energies and the knowledge of their emission processes, as reported here for the high-$z$, Compton-dominated FSRQ PKS 2123$-$463.

\begin{table*}
\footnotesize
\begin{center}
\caption{Model parameters for the SED shown in Fig.~\ref{2123sed}.\label{table_fit}}
\begin{tabular}{lcc}
\hline
Redshift & 	$z$		& 1.46	  \\
Bulk Lorentz Factor & $\Gamma$	& 20	  \\
Doppler Factor & $\delta_{\rm D}$       & 20    \\
Magnetic Field & $B$         & 0.8 G   \\
Variability Timescale & $t_v$       & 1.7$\times$$10^5$ s \\
Comoving radius of blob & R$^{\prime}_b$ & 4.1$\times$10$^{16}$ cm \\
Jet Height & $r$ & 1.0$\times10^{18}$\ cm \\
Low-Energy Electron Spectral Index & $p_1$       & 2.0     \\
High-Energy Electron Spectral Index  & $p_2$       & 3.8	 \\
Minimum Electron Lorentz Factor & $\gamma^{\prime}_{min}$  & $4.0$ \\
Break Electron Lorentz Factor & $\gamma^{\prime}_{brk}$ & 1.0$\times10^3$ \\
Maximum Electron Lorentz Factor & $\gamma^{\prime}_{max}$  & 1.0$\times10^5$ \\
Disc Luminosity & $L_{disc}$ & 1.8$\times10^{46}$\ erg s$^{-1}$ \\
Black Hole Mass & M$_{BH}$   & $2\times10^{9}$ M$_{\odot}$ \\
Accretion efficiency & $\eta$ & 1/12 \\
Gravitational Radius & R$_{g}$ & 1.76$\times10^{14}$ cm \\ 
Inner Disc Radius & R$_{disc\_ISCO}$ & 6 R$_g$ \\
Outer Disc Radius & R$_{disc\_max}$ & 10$^{4}$ R$_g$ \\
Dust Torus luminosity & $L_{dust}$ & 1.5$\times10^{45}$\ erg s$^{-1}$ \\
Dust Torus temperature & $T_{dust}$ & 1.7$\times10^3$\ K \\
Dust Torus radius & $R_{dust}$ & 3.2$\times10^{18}$\ cm \\
Jet Power in Magnetic Field & $P_{j,B}$ & 3.3$\times10^{45}$\ erg s$^{-1}$ \\
Jet Power in Electrons & $P_{j,par}$ & 1.5$\times10^{45}$\ erg s$^{-1}$ \\
\hline
\end{tabular}
\end{center}
\end{table*}

\section*{Acknowledgments}

The {\em Fermi} LAT Collaboration acknowledges generous ongoing support from a
number of agencies and institutes that have supported both the development and
the operation of the LAT as well as scientific data analysis. These include
the National Aeronautics and Space Administration and the Department of Energy
in the United States, the Commissariat \`a l'Energie Atomique and the Centre
National de la Recherche Scientifique / Institut National de Physique
Nucl\'eaire et de Physique des Particules in France, the Agenzia Spaziale
Italiana and the Istituto Nazionale di Fisica Nucleare in Italy, the Ministry
of Education, Culture, Sports, Science and Technology (MEXT), High Energy
Accelerator Research Organization (KEK) and Japan Aerospace Exploration Agency
(JAXA) in Japan, and the K.~A.~Wallenberg Foundation, the Swedish Research
Council and the Swedish National Space Board in Sweden. Additional support for
science analysis during the operations phase is gratefully acknowledged from
the Istituto Nazionale di Astrofisica in Italy and the Centre National
d'\'Etudes Spatiales in France. 

Part of the funding for GROND (both hardware as well as personnel) was generously granted from the Leibniz-Prize to Prof.~G.~Hasinger (DFG  grant
HA~1850/28-1). We thank the Swift team for making these observations possible,
the duty scientists, and science planners. DAK acknowledges support by DFG
grant Kl 766/16-1, and is grateful for travel funding support through
MPE. The Australia Telescope Compact Array is part of the Australia
Telescope National Facility which is funded by the Commonwealth of Australia for operation as a National Facility managed by CSIRO. This
research was funded in part by NASA through Fermi Guest Investigator grant NNH09ZDA001N (proposal number 31263). This research was
supported by an appointment to the NASA Postdoctoral Program at the Goddard Space Flight Center, administered by Oak Ridge Associated
Universities through a contract with NASA. We thank Silvia Rain\'o and the anonymous referee for useful comments and suggestions.

\end{document}